\documentclass[
showpacs,
floatfix,
aps,
prl,
amsmath,
twocolumn,
tightenlines
groupaddress,
]{revtex4}
\usepackage{amsmath,amssymb}
\usepackage{amscd, latexsym}
\usepackage{mathrsfs}
\usepackage{graphicx} 
\usepackage{epstopdf} 
\usepackage{cancel} 
\usepackage{amsfonts}
\usepackage{exscale}
\usepackage{dcolumn}
\usepackage{bm}
\usepackage{color}
\usepackage{comment}

\newcommand{\be}{\begin{equation}}
\newcommand{\ee}{\end{equation}}
\newcommand{\bea}{\begin{eqnarray}}
\newcommand{\eea}{\end{eqnarray}}

\begin{document}

\title{Qubit relaxation from evanescent-wave Johnson noise}

\author{Luke S. Langsjoen, Amrit Poudel, Maxim G. Vavilov, and Robert Joynt}

\affiliation{Department of Physics, University of Wisconsin, Madison, Wisconsin 53706, USA}

\begin{abstract}
In many quantum computer architectures, the qubits are in close proximity to metallic device elements.  
Metals have a high density of photon modes, and the fields spill out of the bulk metal because of 
the evanescent-wave component.  Thus thermal and quantum electromagnetic Johnson-type noise from metallic 
device elements can decohere nearby qubits.  In this paper we use quantum electrodynamics to
compute the strength of this evanescent-wave Johnson noise as a function of distance $z$ from a metallic half-space.  Previous treatments have shown unphysical 
divergences at $z=0$.  We remedy this by using a proper non-local dielectric function.  Decoherence 
rates of local qubits are proportional to the magnitude of electric or magnetic correlation functions 
evaluated at the qubit position. We present formulas for the decoherence rates.  These formulas serve as an
important constraint on future device architectures.
\end{abstract}


\maketitle

Qubits with long relaxation times are necessary for quantum computation. Most
such devices are controlled electrically. This creates a control -- isolation
dilemma: connections from the outside world are what make the devices useful,
but they are also sources of decoherence.  In particular, one may wish to place 
charge or spin qubits close to metallic device elements used to confine or control 
the qubits.  

The relaxation of a charge or spin qubit can be induced by the thermal and quantum 
fluctuations of electromagnetic fields.  The fluctuations of the electromagnetic 
fields are greatly enhanced in the vicinity of conductors
because of the evanescent waves \cite{lifshitz, rytov, pendry, volokitin, joulain}. This evanescent-wave Johnson 
noise (EWJN) has been shown, both theoretically \cite{Henkel1999}
and experimentally \cite{cornell}, to be an important source of decoherence for
atomic qubits near the metallic walls of a trap. In this work we investigate
the effects of metallic device elements in solid-state qubit architectures.
Similar investigations have been carried out previously using lumped --
circuit calculations of Johnson noise \cite{erlingsson, marquardt2009,
Wilhelm2010}. Here we do the noise calculations taking into account the
detailed spatial dependence of the fields and the important effects of
non-local corrections to the electromagnetic response functions \cite{fordweber, Henkel2006}. 

We pause here to mention that the EWJN originates from properties of the metal near its surface.  The expressions for the electromagnetic field fluctuations presented in this paper are for a conducting half space.  However, we have also derived expressions for the strength of EWJN in the vicinity of a conducting slab of finite thickness.  In this case, the electromagnetic fluctuations are independent of the thickness of the slab as long as this thickness is significantly larger than the skin depth of the metal.  As such, we anticipate that EWJN may be alternatively interpreted as arising from overdamped surface plasmon excititations which exist within a skin depth of the surface of the metal.\cite{joulain}

To describe the decoherence of qubits resulting from the evanescent
electromagnetic fields surrounding a conducting gate, it is necessary to
compute spectral densities of the electric and magnetic field fluctuations.
Fermi's Golden Rule and the fluctuation-dissipation
theorem imply that the relaxation rate $1/T_{1}$ of a (charge or spin,
respectively) qubit transition of a particular frequency will be proportional to the 
spectral densities at that frequency. Specifically,  the relaxation time $T_{1}$ of a charge qubit with dipole moment $\vec{d}$ pointing in the $i$th direction at position $\vec{r}$ and level separation $\omega_{Z}$ will be given by
\begin{equation}
\label{T1}
\frac{1}{T_1}=\frac{d^2}{\hbar^2} \chi^E_{ii}(\vec{r}, \vec{r}, \omega_Z) \coth\left(\frac{\hbar\omega_Z}{2k_{B}T}\right), 
\end{equation}
and $T_{1}$ of a spin qubit with magnetic dipole moment $\vec{\mu}$ in the $i$th direction at position $\vec{r}$ and level separation $\omega_{Z}$ will be given by%
\begin{equation}
\label{T2}
\frac{1}{T_1}=\frac{\mu^2}{\hbar^2} \chi^B_{ii}(\vec{r}, \vec{r}, \omega_Z) \coth\left(\frac{\hbar\omega_Z}{2k_{B}T}\right) ,
\end{equation}
where $ \chi^{E,(B)}_{ii}(\vec{r}, \vec{r}, \omega_Z)$ are the electric and magnetic spectral densities, respectively. We deal first with fields that have been averaged over distances of the order
of $a$ and $l$, where $a$ is an interatomic distance and $l$ is the mean free
path in the electrode. As a result the dielectric function $\epsilon(\vec
{r},\omega)$ is a local function of space. This approximation breaks down in
the near vicinity of the conducting surface, and the influence of a nonlocal
dielectric response on the correlation function is addressed below. \ We work
in a gauge where the scalar potential $\phi=0$, so that for harmonic fields we
have $\vec{E}=i\omega\vec{A}/c.$ 

The spectral densities can be shown \cite{LLSP2, joulain} to be directly related to the imaginary part of the equilibrium retarded photon Green's function by the relations
\begin{subequations}
\begin {align}
\chi^E_{ij}(\vec{r}, \vec{r}\tiny~', \omega) &= \frac{\omega^2}{\epsilon_0 c^2} \operatorname{Im}D_{ij}\left(  \vec{r},\vec{r}\tiny~'\normalsize,\omega\right) \\
\chi^B_{ij}(\vec{r}, \vec{r}\tiny~', \omega) &= \frac{1}{\epsilon_0 c^2}\varepsilon_{ikm}\varepsilon_{jnp}\partial_{k}\partial_{n}\operatorname{Im}D_{mp}\left(  \vec{r},\vec{r}\tiny~',\omega \right)
\end{align}
\end{subequations}
where $i,j$ are Cartesian indices that run over $x,y,z$, and $D_{ij}$ satisfies
\begin{align}
&\left[  -\delta_{ij}\left(  \nabla^{2}+\frac{\omega^{2}\epsilon\left(
\vec{r},\omega\right)  }{c^{2}}\right)  +\partial_{i}\partial_{j}\right]
D_{ik}\left(  \vec{r},\vec{r}\tiny~'\normalsize\right)\nonumber \\&  =-4\pi\hslash~\delta^{3}\left(
\vec{r}-\vec{r}\tiny~'\normalsize\right)  \delta_{jk}.\label{eq:pde}%
\end{align}
The geometry of a particular problem is expressed through the function $\epsilon\left(
\vec{r},\omega\right)$.  Equations (\ref{T1}) and (\ref{T2}) assume the charge or spin qubit can be adequately approximated as a point dipole.  The effect of a qubit with an extended spatial distribution will be considered in future work. The task of computing $D_{ij}$ in a particular geometry is, in general, a
complicated problem in electrodynamics. \  In this paper we shall limit
ourselves to the situation where the separation of the qubit from the metal
surface is much less than any radius of curvature of the surface so that the
surface can be thought of as flat.

Let $z$ be the distance from the surface. \ The result for the spectral
density of the electric field for local electrodynamics has been obtained by
Henkel et al \cite{Henkel1999}:
\begin{subequations}
\label{exexezez}
\begin{align}
\label{exex}
 \chi^E_{xx}(z, z, \omega) &=\frac{\hbar}{2\epsilon_0} \text{Re}\int_{0}^{\infty}\frac{pdp}{q}e^{2iqz}  \nonumber \\
&\times   \left(  \frac{\omega^2}{c^2}r_{s}(p)-q^2r_{p}(p)\right)
\end{align}
\begin{align}
\label{ezez}
& \chi^E_{zz}(z, z, \omega) =\frac{\hbar}{\epsilon_0} \text{Re}\int_{0}^{\infty}\frac{p^3}{q}dp e^{2iqz}r_{p}(p)\,, 
\end{align}
\end{subequations}
where $q=\sqrt{\omega^2/c^2-p^2}$ for $p\leq\omega^2/c^2$ and $q=i\sqrt{p^2-\omega^2/c^2}$ for
$p>\omega^2/c^2$ is the $z$-component of the wavevector, and $p$ is the transverse component.  Our notation follows that of Ford and Weber \cite{fordweber}.
\begin{subequations}
\be 
r_{s}\left(  p\right)  =\frac{  q_1-\sqrt{\omega^2\epsilon/c^2-p^{2}}}{q_1+\sqrt{\omega^2\epsilon/c^2-p^{2}}}
\ee
 and
\be
r_{p}\left(p\right)  =\frac{\epsilon q_1-\sqrt{\omega^2\epsilon/c^2-p^{2}}}{\epsilon q_1+\sqrt{\omega^2\epsilon/c^2-p^{2}}}
\ee
\end{subequations}
are the Fresnel reflection coefficients.  The corresponding expressions for the spectral densities of the magnetic field are identical to Eqs. (\ref{exexezez}) if we multiply by $1/c^2$ and make the replacement $r_s\leftrightarrow r_p$.
 
We are interested in separations that are sufficiently small so that retardation, and hence, radiation of the electromagnetic field may be neglected. This is known as the quasistatic approximation, and it is formally employed by taking the limit $c\rightarrow\infty$.  This results in the greatly simplified expressions
\begin{subequations}
\label{exexezezqs}
\begin{align}
\label{exexqs}
&\chi^E_{zz}(z, z, \omega) = 2 \chi^E_{xx}(z, z, \omega) =\frac{\hbar }{8\epsilon_0  z^3}\text{Im}\frac{\epsilon-1}{\epsilon+1}~, \\
\label{bxbxqs}
&\chi^B_{zz}(z, z, \omega) = 2 \chi^B_{xx}(z, z, \omega) =\frac{\hbar \omega^2}{8\epsilon_0c^4 z} \text{Im}(\epsilon-1) \,.
\end{align}
\end{subequations}

 Employing this approximation has eliminated the functional difference between the 
transverse and longitudinal components of the field fluctuations. The quasistatic expressions 
differ from the exact values by less than $1\%$ when $z<\delta/10$, where $\delta$ is the skin 
depth of the metal. For copper near absolute zero, $\delta\sim3\mu m$. Equations (\ref{exexezezqs}) 
diverge as $z\rightarrow0$, but this divergence is not
physical: it is an artifact of treating the dielectric function as local at
distances comparable to the interatomic spacing.
In any differential equation satisfied by the spatial Fourier components of
$D_{ij}$, a local dielectric function will be independent of the wavevector
while a nonlocal one will have a nontrivial wavevector dependence. It is
conventional to represent the spatial Fourier components of the nonlocal dielectric function as a tensor
quantity in the form
\begin{equation}
\epsilon_{ij}(\vec{k},\omega)=\epsilon_{l}(k,\omega)\frac{k_{i}k_{j}}{k^{2}}+\epsilon_{t}(k,\omega)\left(
\delta_{ij}-\frac{k_{i}k_{j}}{k^{2}}\right)  ,\label{nonlocalepsilon}%
\end{equation}
where we have separated the function into its longitudinal $\epsilon_{l}$ and
transverse $\epsilon_{t}$ components. 


\begin{figure}
\includegraphics[width = 1.0 \columnwidth] {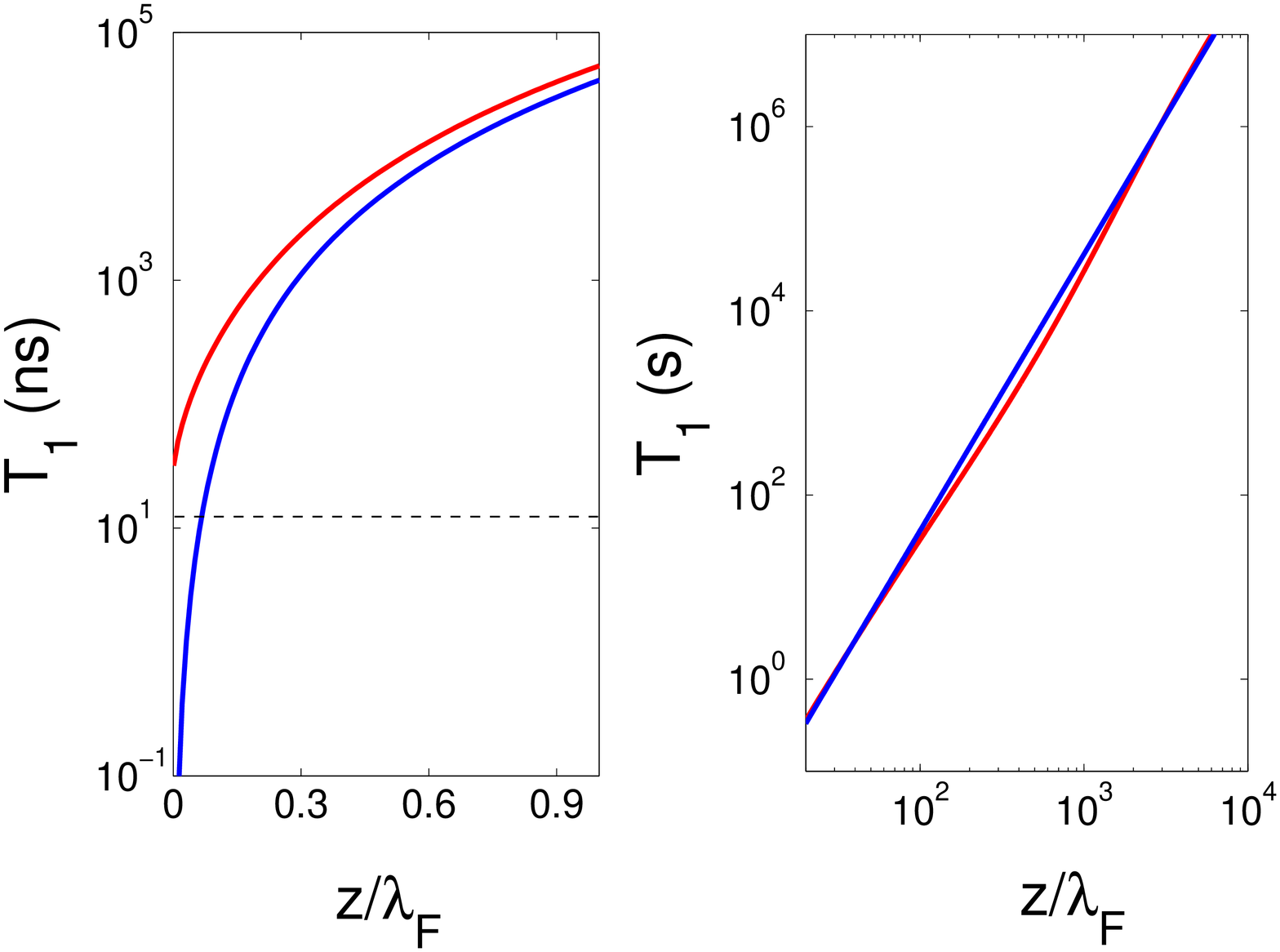}
\caption {(Color online) Plot of the $T_1$ time of a charge qubit computed from Eq. (1) using the local approximation ({\color{blue}blue}) and 
the full nonlocal theory ({\color{red}red}).  We used the values 
$E_F=7eV$, $\omega=6\pi \times10^8s^{-1}$, $\nu=6\pi\times10^{12}s^{-1}$, and $\omega_p=1.6\times10^{16}s^{-1}$, appropriate
for a copper surface and a device operating in the GHz range. 
The dipole moment is taken as $d = |e| a_B$, where $|e|$ is minus the charge on the electron and $a_B$ is the Bohr radius.
These results are for zero temperature.  The dashed horizontal line in the left figure represents the strength of the electric field fluctuations inside the bulk of a uniform metal.}       
\end{figure}

\begin{figure} [t]
\includegraphics[width = 1.0 \columnwidth] {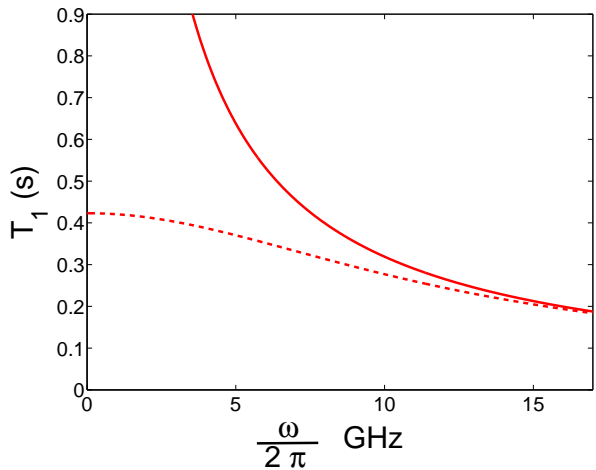}
\caption { (Color online) Plot of the $T_1$ time $vs.$ frequency $\omega$ of a charge qubit 
computed from Eq. (1) at different temperatures  (solid red at 0 K and dashed red at 2 K) at fixed 
$z = 10 \lambda_F$. $E_F=7eV$, $\nu=6\pi\times10^{12}s^{-1}$, $\omega_p=1.6\times10^{16}s^{-1}$, and
$d = |e| a_B$, as in Fig. 1.}
\end{figure}

Nonlocality in the dielectric function changes the reflection coefficients. In the quasistatic
approximation $r_{p}$ is \cite{fordweber}
\begin{subequations}
\label{rnonlocal}
\begin{equation}
r_{p}=\frac{ 1-\epsilon_{0}\dfrac{2p}{\pi}\displaystyle\int_{0}^{\infty}d\kappa\dfrac{1}{k^{2}\epsilon_{l}(k,\omega)}}{  1+\epsilon_{0}\dfrac{2p}{\pi}\displaystyle\int_{0}^{\infty}d\kappa\dfrac{1}{k^{2}\epsilon_{l}(k,\omega)}}
\label{rpnonlocal}
\end{equation}
and $r_s$ becomes
\begin{equation}
r_s=\frac{\omega^2}{4p^2c^2}\left(\frac{4p^3}{\pi\epsilon_0}\int_0^\infty\frac{\epsilon_t(k,\omega)}{k^4}d\kappa-1\right)
\end{equation}
\end{subequations}

to leading nonvanishing order in the quasistatic approximation. Here $k^2=p^2+\kappa^2$,
\begin{subequations}
\begin{align}
\epsilon_{l}(k,\omega)&=1+\frac{3\omega_{p}^{2}}{k^2v_F^2}\frac{(\omega+i\nu)%
f_{l}((\omega+i\nu)/kv_F)}{\omega+i\nu f_{l}((\omega+i\nu)/kv_F)}\\
\epsilon_t(k,\omega)&=1-\frac{\omega_p^2}{\omega(\omega+i\nu)}f_t((\omega+i\nu)/kv_F)
\end{align}
\end{subequations}
\begin{subequations}
\begin{align}
f_{l}(x)&=1-\frac{x}{2}\ln(x+1)/(x-1)\\
f_{t}(x)&=\frac{3}{2}x^2-\frac{3}{4}x(x^2-1)\ln(x+1)/(x-1)
\end{align}
\end{subequations}
$\nu$ is the electron collision frequency, $\omega_{p}=(4\pi ne^{2}/m)^{1/2}$
is the plasma frequency, and $v_{F}$ is the
Fermi velocity. 
Although Eqs. (\ref{exexezez}) are derived assuming locality, it is a convenient fact \cite{fordweber} that these equations are also valid in the nonlocal regime, as long as the nonlocal form of the Fresnel coefficients Eqs. (\ref{rnonlocal}) are used.  

In Figure 1, we present the zero-temperature results for the relaxation time $T_1$ from EWJN for a qubit with an electric dipole moment of magnitude $|e|a_B$, where $a_B$ is the Bohr radius. Both the local and nonlocal results are shown. It is seen that the correct nonlocal dielectric function eliminates the unphysical divergence of $1/T_1$ at $z=0$. For separations $z\sim\lambda_F$, the differences are very significant, while for $z\gtrsim10\lambda_F$, the local and nonlocal results nearly coincide. The Fermi wavelength is less than a nanometer, so the transition from local to nonlocal behavior occurs well within the quasistatic regime. It is interesting to note that for the electric field fluctuations there is a crossover region where the nonlocal result becomes slightly larger than the local result in the range $30\lambda_F<z<3000\lambda_F$  (see Fig. 1), in alignment with the results of Volokitin et al \cite{volokitin} who showed an enhancement of the nonlocal result above the local result. We see that at $T=0$ and GHz operations, $T_1$ from spontaneous emission is of the order of seconds at separations $z\sim30\lambda_F$. These results are directly applicable to atomic qubits, but the rate $1/T_1$ is proportional to the square of the dipole moment, so rates for other charge qubits are easily deduced.  Figure 2 shows that frequency dependence of $T_1$ falls off slowly at higher frequencies, but can be very strong at low temperatures $T<\hbar\omega/k_B$.

Figure 3 gives the analogous results for magnetic EWJN on a spin qubit with a magnetic dipole moment of 1 Bohr magneton. Nonlocal corrections are somewhat stronger for this case, and persist to larger distances. Interestingly, the falloff with distance of $T_1$ is slower for magnetic EWJN than for electric EWJN. However, magnetic relaxation times are typically somewhat larger than electric relaxation times. The crossover of the local and nonlocal results is not present in the magnetic case. Figure 4 shows that the frequency and temperature dependence of magnetic EWJN is similar to the electric EWJN shown in Fig. 2. Brief mention should be made of the dip that is observed as $\omega\rightarrow0$ in Fig. 4. The reflection coefficients $r_s$ and $r_p$ both contribute to the magnetic field fluctuations to the same order in $\omega/c$ in the quasistatic approximation.  This contrasts with the electric case, where only $r_p$ contributes to leading order in $\omega/c$.  The dip is a result of competition between the contributions of $r_s$ and $r_p$ to the field fluctuations.  The $r_s$ term in $\chi^B_{ii}$ is linear in $\omega$ as $\omega\rightarrow0$ and negative, while the $r_p$ term is cubic in $\omega$ as $\omega\rightarrow0$ and positive. We should also mention that for extremely small $\omega$ the factor of $\coth(\hbar\omega/2k_BT)$ cancels out all $\omega$ dependence and the dip flattens out as $\omega\rightarrow0$ (not observable in the resolution of Fig. 4).

The limit of the nonlocal quasistatic field fluctuations as $z\rightarrow 0$ should be of the same order of magnitude as the value of these fluctuations inside the metal.  To check this, we calculate the electromagnetic Green's function inside the bulk of a uniform metal using a nonlocal dielectric function.  The result is
\begin{align}
D_{ij}(\vec{k},\omega)&=\frac{4\pi\hbar}{\omega^2\epsilon_t/c^2-k^2}\nonumber\\ &\times\left(\delta_{ij}-\frac{c^2k_ik_j}{\omega^2\epsilon_l}+\frac{k_ik_j}{k^2\epsilon_l}(\epsilon_t-\epsilon_l)\right)
\end{align}
\be
\label{Dijuniformmetal}
D_{ij}(\vec{r}-\vec{r}\tiny~'\normalsize,\omega)=\frac{1}{(2\pi)^3}\int d^3\vec{k}e^{i\vec{k}\cdot(\vec{r}-\vec{r}\tiny~'\normalsize)}D_{ij}(\vec{k},\omega)
\ee

\begin{figure} [t]
\includegraphics[width = 1.0 \columnwidth] {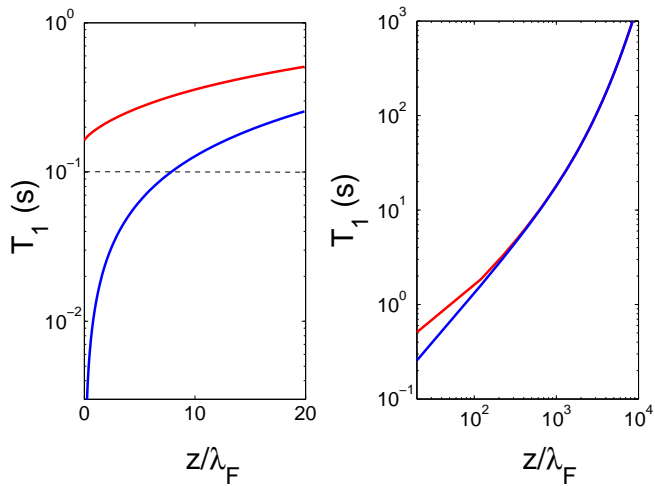}
\caption { (Color online) Plot of $T_1$ time for a spin qubit at zero temperature computed from Eq. (2) in the local approximation ({\color{blue}blue}) 
and the full nonlocal theory ({\color{red}red}).   $E_F=7eV$, $\omega=6\pi \times10^8s^{-1}$, $\nu=6\pi\times10^{12}s^{-1}$, and $\omega_p=1.6\times10^{16}s^{-1}$, appropriate
for a copper surface and a device operating in the GHz range. We have taken $\mu = \mu_B$, appropriate for a single electron.  The rate $1/T_1$
is proportional to the square of $\mu$, so rates for other local magnetic qubits can be easily deduced.  The dashed horizontal line in the left figure represents the strength of the magnetic field fluctuations inside the bulk of a uniform metal.}
\end{figure}

\begin{figure}
\includegraphics[width = 1.0 \columnwidth] {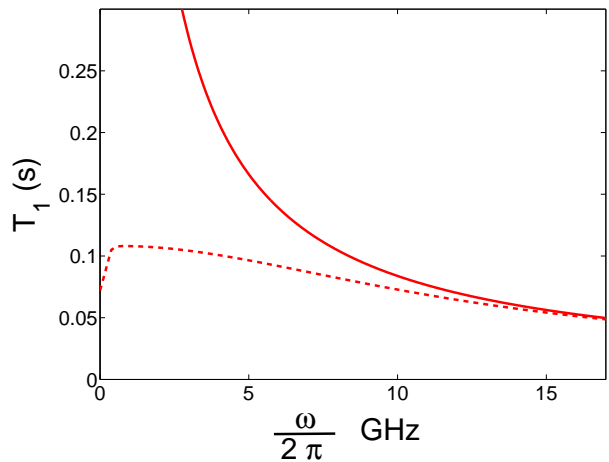}
\caption { (Color online) Plot of the $T_1$ time $vs.$ frequency of  spin qubit computed at different temperatures (solid red at 0 K and dashed red at 2 K) at fixed $z = 10 \lambda_F$.
$E_F=7eV$, $\nu=6\pi\times10^{12}s^{-1}$, and $\omega_p=1.6\times10^{16}s^{-1}$, appropriate
for a copper surface. $\mu = \mu_B$ as in Fig. 3.}
\end{figure}

Numerical evaluation of (\ref{Dijuniformmetal}) when $\vec{r}=\vec{r}\tiny~'\normalsize$ gives $D_{xx}= D_{zz}\sim3.2\times10^{-15} Js/m$. An evaluation of the Green's function outside the metal in the nonlocal quasistatic regime for $z\rightarrow 0$ gives $D_{xx}\sim1.32\times10^{-15} Js/m$ and $D_{zz}\sim2.6\times10^{-15} Js/m$, slightly less than in the bulk of a uniform metal, as expected.

We see that there are three relevant distance regimes.  For $z<30\lambda_F$, a quasistatic approximation to Henkel's results (\ref{exexezez}) using the \textit{nonlocal} expression for $r_p$ will accurately describe the field fluctuations.  For intermediate distances $30\lambda_F<z<\delta/10$, there is a slight enhancement in the electric field fluctuations from the nonlocal expression for $r_p$ compared to the local expression. For distances $z>\delta/10$, quasistatic local forms (\ref{exexezez}) will accurately describe the field fluctuations. 

The density of photon states in a metal is very high owing to the large polarizability.  For blackbody radiation, this
high density of states does not matter, since total internal reflection reduces the outgoing radiation
flux to its universal Stefan-Boltzmann value.  In contrast, the evanescent waves are strongly enhanced
and the resultant electromagnetic noise just outside the surface can be intense.  This is a concern for 
quantum devices operating close to metallic objects.  This paper has concentrated on the frequency, 
temperature, and distance dependence of the noise, and on the effects of assuming a local dielectric function.
We conclude that the effect is significant for charge qubits with large dipole moments
such as double quantum dots.  The EWJN relaxation may be the limiting decoherence effect in designs which involve close proximity to bulk metals.  For magnetic qubits the effects are smaller.  We found that nonlocal effects are 
very important at short distances - indeed, local calculations can produce spurious divergences.  At
distances large compared to the Fermi wavelength of the metal, local approximations work well.

We have not considered extended qubits for which the off-diagonal function $D_{ii}(\vec{r},\vec{r}\tiny~'\normalsize)$ 
at $\vec{r} \neq \vec{r}\tiny~'\normalsize$ is required.  We have treated only relaxational decoherence and have
ignored the possibility of dephasing.  These effects are important for many real devices and can
be calculated using similar methods.         

We thank M.A. Eriksson, S.M. Girvin, R. McDermott, and F. Wellstood for useful discussions.  This work was supported by ARO and LPS grant no. W911NF-11-1-0030.

\end{document}